\documentstyle[12pt]{article}	
\begin{document}

\setcounter{page}{1}

\title{\small\centerline{August 1993 \hfill DOE-ER\,40757-023 / CPP-93-23}
\medskip
{\LARGE\bf Equivalence theorem\\[3mm] and dynamical symmetry
breaking}\bigskip}

\author{Palash B. Pal\\
\normalsize \em Center for Particle Physics, University of Texas,
              Austin, TX 78712, USA}

\date{}
\maketitle

\begin{abstract} \normalsize\noindent
The equivalence theorem, between the longitudinal gauge bosons and the
states eaten up by them in the process of symmetry breaking, is shown
to be valid in a class of models where the details of dynamical
symmetry breaking makes it obscure.
\end{abstract}
\bigskip\bigskip

%

In a recent paper \cite{DoTa93}, Donoghue and Tandean have made an
intriguing point regarding the validity of the equivalence theorem
\cite{CLT74,BQT77,ChGa85,Wil88,BaSc90,Vel90} in a class of gauge models
which exhibit dynamical symmetry breaking.  As a paradigm, they
considered the pedagogical model \cite{Wei79,Sus79} where the
electroweak gauge symmetry is dynamically broken by quark condensates
which are also responsible for the chiral symmetry breaking:
	\begin{eqnarray}
\langle \overline u_L u_R \rangle =
\langle \overline d_L d_R \rangle \neq 0 \,.
	\end{eqnarray}
The pions, $\pi^\pm\pi^0$, which would have been the Goldstone bosons
for the chiral symmetry breaking $SU(2)_L\times SU(2)_R \to SU(2)_V$,
are eaten up by the electroweak gauge bosons in this model in absence
of any fundamental Higgs bosons. Donoghue and Tandean \cite{DoTa93}
then calculate the amplitude of the process $e^+e^-\to Z\gamma$ at
energies larger than $M_Z$ mediated by the triangle diagram. Both
quarks and leptons can appear in the triangle, and from the condition
of vanishing of gauge anomalies, they obtain that the amplitude
vanishes. Since $\pi^0$ constitutes the longitudinal part of
the $Z$ in this model, they then compare it with the process
$e^+e^-\to \pi^0\gamma$, mediated through the triangle diagram.
However, since the pion consists of quarks but no leptons, only quarks
circulate in the loop now, and therefore the amplitude does not
vanish.  This, they claim, is a violation of the equivalence theorem.
The purpose of this article is to examine this claim.

In Higgs models of symmetry breaking, one can calculate the couplings
of the unphysical Higgs bosons (which are eaten up by the gauge bosons)
from those of the gauge bosons without having to make any assumption
about the Higgs content of the model.  This can be done by demanding
that, if one uses the $R_\xi$ gauges to calculate any amplitude, the
$\xi$-dependent poles must cancel \cite{FLS72,CBLP84}.  Thus, for two
fermions $a$ and $b$ which are represented by their field operators
$\psi_a$ and $\psi_b$, if the coupling to any gauge boson $V$ is given
by
	\begin{eqnarray}
\overline \psi_a \gamma^\mu \left( {\cal G}_{ab} + {\cal G}'_{ab}
\gamma_5 \right) \psi_b V_\mu \,,
\label{Vcoupling}
	\end{eqnarray}
this requirement demands that the corresponding unphysical Higgs
boson, $S$, will have the coupling
	\begin{eqnarray}
{1\over M_V} \overline \psi_a \left[ (m_a - m_b) {\cal G}_{ab} +
(m_a + m_b) {\cal G}'_{ab} \gamma_5 \right] \psi_b S \,,
\label{Scoupling}
	\end{eqnarray}
where $M_V$ is the mass of the gauge boson $V$ after symmetry
breaking.  From this requirement alone, one can verify the equivalence
theorem for any amplitude.

However, the equivalence theorem is more deep-rooted than the Higgs
mechanism of symmetry breaking for the following reason. In any gauge
symmetry breaking, some gauge bosons obtain masses $M_V$ whose exact
values depend on some parameters of the theory.  For any
$M_V \neq 0$, the longitudinal components of the massive gauge bosons
are physical states.  The ``Nambu-Goldstone'' modes,  the states
absorbed by the gauge bosons, are unphysical. On the other hand, for
$M_V=0$,  the Nambu-Goldstone modes are physical states but the
longitudinal components are not, since the symmetry is not broken.
The equivalence theorem then merely states that all observables are
continuous in the limit $M_V\to 0$. In other words, in that limit, the
amplitudes with any process with longitudinal component of a vector
boson is the same (apart from a phase maybe) with the amplitudes for
the corresponding processes where the longitudinal gauge bosons are
replaced by the states that are eaten up by them in the process of
symmetry breaking.  Stated this way, the equivalence theorem seems to
be a statement of continuity of certain parameters of the theory, and
hence is expected to be valid for any model with symmetry breaking. In
light of this, the claim of Donoghue and Tandean \cite{DoTa93} is
indeed surprising.

To examine their claim, let us use their paradigm of QCD condensates
breaking the electroweak gauge symmetry \cite{Wei79,Sus79}, and for
the sake of definiteness, let us talk about processes involving $Z$
bosons only.  Obviously, one can make similar arguments for
processes involving $W$ bosons.  Since the $Z$-boson does not
provide any flavor changing neutral current with the standard model
fermions, the indices $a$ and $b$ in Eq.\ (\ref{Vcoupling}) have to be
equal, and we denote the couplings  ${\cal G}$ and ${\cal G}'$ in this
case with a single index.  From Eq.\ (\ref{Scoupling}) now, we
see that proving the equivalence theorem is tantamount to proving that
the coupling of the $\pi^0$ to the fermion field $\psi_a$ is given by
	\begin{eqnarray}
{2m_a \over M_Z}\, {\cal G}'_a \; \overline \psi_a \gamma_5 \psi_a
\pi^0  \,.
\label{zcoupling}
	\end{eqnarray}

	\begin{figure}
\large
\begin{center}
\begin{picture}(85,36)(-5,32)
\thicklines
\put(15,50){\line(-1,1){15}}     \put(7,58){\vector(1,-1){2}}
\put(4,62){$a(p)$}			
\put(15,50){\line(-1,-1){15}}      \put(7,42){\vector(-1,-1){2}}
\put(4,35){$a(p')$} 			
\multiput(16,50)(4,0){10}{\oval(2,2)[t]}
\multiput(18,50)(4,0){10}{\oval(2,2)[b]}
\put(35,48){\makebox(0,0){\vector(1,0){5}}}
\put(35,54){\makebox(0,0)[b]{$Z^\mu(q)$}}
\put(55,50){\circle*{3}}
\multiput(55,50)(3,0){10}{\line(1,0){2}}
\put(70,52){\makebox(0,0)[b]{$\pi^0(q)$}}
\end{picture}
\end{center}
\caption[]{\sf Tree level diagram giving rise to the pion coupling with
fermions.}
\label{f:coupling}
	\end{figure}
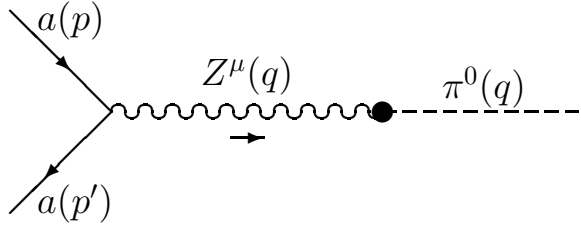
If the fermion $a$ in question is the electron, for example, it might
naively seem that such a coupling with the pion cannot exist since the
pion wave function does not have any electron. However, this is not
true, as can be seen from the diagram of Fig.\ \ref{f:coupling}.  Two
important points need to be made before we calculate this coupling.
First, the intermediate line can only be $Z$. The diagram with
intermediate photon line cannot contribute since the photon couplings
are vectorial, whereas the pion couples only to the axial vector
current through the relation
	\begin{eqnarray}
\left< 0 \left| \overline Q \gamma^\mu \gamma_5 {\tau^I\over 2} Q
\right| \pi^J (q) \right> = f_\pi q^\mu \delta^{IJ} \,,
\label{fpi}
	\end{eqnarray}
where $|0\rangle$ is the hadronic vacuum, $I,J$ indices run over the
adjoint representation of the isospin symmetry SU(2)$_V$, and
	\begin{eqnarray}
Q \equiv \left( \begin{array}{c} u \\  d \end{array} \right) \,.
	\end{eqnarray}
Second, the diagram must be calculated in the unitary gauge where
unphysical degrees of freedom cannot appear in the intermediate state.
Otherwise, the pions can appear even as intermediate states and it
will be impossible to calculate the diagram.
In the unitary gauge, the propagator of the gauge boson is given by
	\begin{eqnarray}
-i D^{\mu\nu} (q) = {-i (g^{\mu\nu} - q^\mu q^\nu/M_Z^2) \over
q^2-M_Z^2} \,.
\label{Zpropag}
	\end{eqnarray}
Thus the effective interaction between the fermions and the pion
derived from this diagram is given by\footnote{The up-quark field $u$
(in italics) is not to be confused with the positive energy spinor
$\bf u$ (in boldface).}
	\begin{eqnarray}
i {\cal L}_{\rm eff} =
{\bf \overline u_{\em a} (p')} \, i \gamma_\mu
\left( {\cal G}_a + {\cal G}'_a \gamma_5
\right) {\bf u_{\em a} (p)} \cdot \left[ -i D^{\mu\nu} (q) \right] \cdot
i\left< 0 \left| J_\nu^{(Z)} \right| \pi^0 (q) \right> \,,
\label{picoup}
	\end{eqnarray}
where $J_\nu^{(Z)}$ is the current that couples to the $Z$ boson:
	\begin{eqnarray}
J_\nu^{(Z)} = -\, {g \over 4\cos \theta_W} \left(
\overline u \gamma_\nu \gamma_5 u - \overline d \gamma_\nu \gamma_5 d
+ \cdots \right) \,,
\label{jz}
	\end{eqnarray}
where $g$ is the weak SU(2) gauge coupling, $\sin\theta_W=e/g$,
and the dots signify vector currents as well as currents of fermions
other than the up and the down quarks which are of no interest for us.
{}From Eqs.\ (\ref{fpi}), (\ref{Zpropag}) and (\ref{jz}), we obtain
	\begin{eqnarray}
D^{\mu\nu} (q) \; \left< 0 \left| J_\nu^{(Z)} \right| \pi^0 (q)
\right> = {q^\mu \over M_Z} \,,
\label{DJ}
	\end{eqnarray}
using the relations for the gauge boson masses obtained in this model,
viz.,
	\begin{eqnarray}
M_W = M_Z \cos \theta_W = {1\over 2} gf_\pi \,.
\label{MWMZ}
	\end{eqnarray}
Putting Eq.\ (\ref{DJ}) in Eq.\ (\ref{picoup}),
it is straightforward to show that the amplitude is
	\begin{eqnarray}
{2m_a \over M_Z} \, {\cal G}'_a {\bf  \overline u_{\em a} (p')}
\gamma_5 {\bf u_{\em a} (p)} \,,
\label{finalcoup}
	\end{eqnarray}
where we have used the spinor definitions
	\begin{eqnarray}
\rlap/p {\bf u (p)} = m {\bf u (p)} \,, \qquad
 {\bf\overline u (p)} \rlap/p = m {\bf\overline u (p)} \,.
\label{spinor}
	\end{eqnarray}
Obviously, Eq.\ (\ref{finalcoup}) is equivalent to the interaction of
Eq.\ (\ref{zcoupling}).  Since this coupling is obtained in this
model, it is now easy to verify that the equivalence theorem is valid
for any  amplitude, as we argued before.

	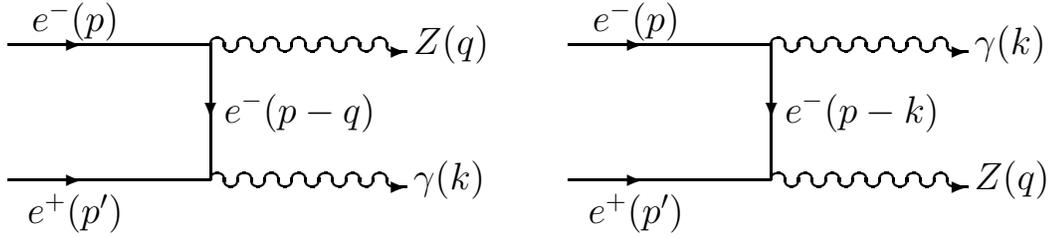
\begin{figure}
\large
\begin{center}
		\begin{picture}(70,30)(0,35)
\thicklines
\put(5,60){\line(1,0){30}}	\put(15,60){\vector(1,0){1}}
\put(15,62){\makebox(0,0)[b]{$e^-(p)$}}
\put(5,40){\line(1,0){30}}	\put(15,40){\vector(1,0){1}}
\put(15,38){\makebox(0,0)[t]{$e^+(p')$}}
\put(35,40){\line(0,1){20}}	\put(35,50){\vector(0,-1){1}}
\put(37,50){\makebox(0,0)[l]{$e^-(p-q)$}}
\multiput(36,60)(4,0){7}{\oval(2,2)[t]}
\multiput(38,60)(4,0){6}{\oval(2,2)[b]}
\put(62,60){\oval(2,2)[bl]}
\put(62,59){\vector(1,0){2}}
\put(65,60){\makebox(0,0)[l]{$Z(q)$}}
\multiput(36,40)(4,0){7}{\oval(2,2)[t]}
\multiput(38,40)(4,0){6}{\oval(2,2)[b]}
\put(62,40){\oval(2,2)[bl]}
\put(62,39){\vector(1,0){2}}
\put(65,40){\makebox(0,0)[l]{$\gamma(k)$}}
\end{picture}
				\qquad
		\begin{picture}(70,30)(0,35)
\thicklines
\put(5,60){\line(1,0){30}}	\put(15,60){\vector(1,0){1}}
\put(15,62){\makebox(0,0)[b]{$e^-(p)$}}
\put(5,40){\line(1,0){30}}	\put(15,40){\vector(1,0){1}}
\put(15,38){\makebox(0,0)[t]{$e^+(p')$}}
\put(35,40){\line(0,1){20}}	\put(35,50){\vector(0,-1){1}}
\put(37,50){\makebox(0,0)[l]{$e^-(p-k)$}}
\multiput(36,60)(4,0){7}{\oval(2,2)[t]}
\multiput(38,60)(4,0){6}{\oval(2,2)[b]}
\put(62,60){\oval(2,2)[bl]}
\put(62,59){\vector(1,0){2}}
\put(65,60){\makebox(0,0)[l]{$\gamma(k)$}}
\multiput(36,40)(4,0){7}{\oval(2,2)[t]}
\multiput(38,40)(4,0){6}{\oval(2,2)[b]}
\put(62,40){\oval(2,2)[bl]}
\put(62,39){\vector(1,0){2}}
\put(65,40){\makebox(0,0)[l]{$Z(q)$}}
\end{picture}
\end{center}
\caption[]{\sf Tree level diagrams for the process $e^+e^-\to Z\gamma$.}
\label{f:Zgam}
\end{figure}
The skeptic in us may wonder whether our proof is valid for processes
where $Z$ or $\pi^0$ couples to internal fermion lines, given that we
have used the on-shell condition for the spinors, Eq.\ (\ref{spinor}),
to derive Eq.\ (\ref{finalcoup}).\footnote{In fact, one can ask the
same question about Higgs models of symmetry breaking since Eq.\
(\ref{spinor}) is used to derive Eq.\ (\ref{Scoupling}) as well.}  We
put to rest such doubts by explicitly calculating the process
$e^+e^-\to Z\gamma$, which is the process calculated by Donoghue and Tandean
\cite{DoTa93}.  However, we note that the triangle diagrams considered
by them are not the lowest order diagrams for this process. There are
tree diagrams, given in Fig.\ \ref{f:Zgam}, which contribute.  The
amplitude for this diagram, ${\cal A}_Z$, can be written as
	\begin{eqnarray}
i{\cal A}_Z = ie \varepsilon^\mu (q) \epsilon^\nu(k) {\bf \overline v
(p')} \Gamma_{\mu\nu} {\bf u (p)} \,,
\label{AZ}
	\end{eqnarray}
where $\varepsilon$ and $\epsilon$ represent the polarization vectors
of the $Z$ and the photon respectively, and
	\begin{eqnarray}
\Gamma_{\mu\nu} = \gamma_\nu {\rlap/p - \rlap/q + m_e \over (p-q)^2 -
m_e^2} \gamma_\mu \left( {\cal G}_e + {\cal G}'_e \gamma_5 \right) +
\gamma_\mu \left( {\cal G}_e + {\cal G}'_e \gamma_5 \right)
 {\rlap/p - \rlap/k + m_e \over (p-k)^2 - m_e^2}  \gamma_\nu \,.
	\end{eqnarray}

The diagrams for $e^+e^-\to \pi^0\gamma$, on the other hand, are
obtained if the $Z$-boson lines of Fig.\ \ref{f:Zgam} couple to the
pion wavefunction in the manner shown in Fig.\ \ref{f:coupling}.  The
amplitude of such diagrams is given by
	\begin{eqnarray}
i{\cal A}_\pi &=& ie \epsilon^\nu(k) {\bf \overline v
(p')} \Gamma_{\mu\nu} {\bf u (p)}  \cdot
\left[ -i D^{\mu\rho} (q) \right] \cdot
i\left< 0 \left| J_\rho^{(Z)} \right| \pi^0 (q) \right> \nonumber\\
&=& ie \epsilon^\nu(k) {\bf \overline v
(p')} \Gamma_{\mu\nu} {\bf u (p)}  \, {q^\mu \over M_Z} \,,
\label{Api}
	\end{eqnarray}
using Eq.\ (\ref{DJ}) in the last step.  In general, the amplitudes
in Eqs.\ (\ref{AZ}) and (\ref{Api}) are not equal, even in magnitude.
However, if we consider a longitudinal polarized $Z$ boson in Eq.\
(\ref{AZ}) whose 4-momentum is given by $(E,\kappa\hat n)$ for some unit
3-vector $\hat n$, the polarization vector will be given by
$\varepsilon^\mu_{\rm long} (q) \equiv (\kappa,E\hat n)/M_Z$.  In the limit
$M_Z\to 0$ or equivalently $E/ M_Z\to \infty$, since $\kappa\approx
E$, we obtain $\varepsilon^\mu_{\rm long} (q) = q^\mu/M_Z$.  Thus, in
this limit, the amplitudes for $e^+e^-\to Z_{\rm long}\gamma$ and
$e^+e^-\to \pi^0\gamma$ are indeed equal, as seen from Eqs.\
(\ref{AZ}) and (\ref{Api}).  This is the verification of
the equivalence theorem to this order.

It is now easy to see that the proof can be extended to any process,
e.g., with any number of $Z$-bosons in the initial and final states.
 If we have a diagram with an external $Z_{\rm
long}$ line, we obtain a factor $\varepsilon^\mu_{\rm long} (q)$ in
the amplitude.  On the other hand, if we let the $Z$-boson to couple
to the $\pi^0$ wavefunction, we will obtain the factors
$\left[ -i D^{\mu\rho} (q) \right] \cdot
i\left< 0 \left| J_\rho^{(Z)} \right| \pi^0 (q) \right>$, coming from
the $Z$ propagator and the matrix element involving the pion
wavefunction. However, as shown in Eq.\ (\ref{DJ}), this equals
$q^\mu/M_Z$, which is $\varepsilon^\mu_{\rm long} (q)$ in the limit
$M_Z\to 0$.  Thus, equivalence theorem is valid for any process.

	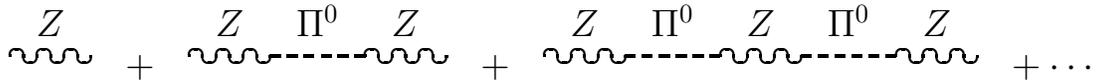
\begin{figure}
\large
\begin{center}
		\begin{picture}(16,10)(-2,47)
\thicklines
\multiput(1,50)(4,0){3}{\oval(2,2)[t]}
\multiput(3,50)(4,0){3}{\oval(2,2)[b]}
\put(6,53){\makebox(0,0)[b]{$Z$}}		
		\end{picture}
$\null + \null$
		\begin{picture}(42,10)(-2,47)
\thicklines
\multiput(1,50)(4,0){3}{\oval(2,2)[t]}
\multiput(3,50)(4,0){3}{\oval(2,2)[b]}
\put(6,53){\makebox(0,0)[b]{$Z$}}		
\multiput(12,50)(2.5,0){6}{\line(1,0){1.5}}
\put(19,53){\makebox(0,0)[b]{$\Pi^0$}}		
\multiput(27,50)(4,0){3}{\oval(2,2)[t]}
\multiput(29,50)(4,0){3}{\oval(2,2)[b]}
\put(32,53){\makebox(0,0)[b]{$Z$}}		
		\end{picture}
$\null + \null$
		\begin{picture}(68,10)(-2,47)
\thicklines
\multiput(1,50)(4,0){3}{\oval(2,2)[t]}
\multiput(3,50)(4,0){3}{\oval(2,2)[b]}
\put(6,53){\makebox(0,0)[b]{$Z$}}		
\multiput(12,50)(2.5,0){6}{\line(1,0){1.5}}
\put(19,53){\makebox(0,0)[b]{$\Pi^0$}}		
\multiput(27,50)(4,0){3}{\oval(2,2)[t]}
\multiput(29,50)(4,0){3}{\oval(2,2)[b]}
\put(32,53){\makebox(0,0)[b]{$Z$}}		
\multiput(38,50)(2.5,0){6}{\line(1,0){1.5}}
\put(45,53){\makebox(0,0)[b]{$\Pi^0$}}		
\multiput(53,50)(4,0){3}{\oval(2,2)[t]}
\multiput(55,50)(4,0){3}{\oval(2,2)[b]}
\put(58,53){\makebox(0,0)[b]{$Z$}}		
		\end{picture}
$\null + \cdots$
		\end{center}
\caption[]{\sf Diagrams responsible for the mass of the $Z$-boson.}
\label{f:Zpole}
\end{figure}
And in fact, it is also easy to see that the proof can be easily
extended to any model of dynamical symmetry breaking.  In general, let
us denote the spin-0 state eaten up by the $Z$-boson by $\Pi^0$. Since
the $Z$ mass is generated by the series of diagrams given in Fig.\
\ref{f:Zpole}, it is easy to see that one requires
	\begin{eqnarray}
\left< 0 \left| J_\nu^{(Z)} \right| \Pi^0 (q)
\right> = - M_Z q_\nu \,,
	\end{eqnarray}
no matter how $M_Z$ is related to the parameters of the unbroken
theory.\footnote{Actually, the right side can have an arbitrary phase,
which will appear as an overall phase of all couplings of $\Pi^0$.
But this phase does not affect any physics, including the equivalence
theorem.}      This is all one needs to verify Eq.\ (\ref{DJ}), and
thereby the equivalence theorem.

Earlier, we said that the equivalence theorem is the statement of
continuity of physical observables in the limit $M_V\to 0$.
Since this limit can be realized as $g\to 0$, it
is obvious that one expects the equivalence at only the lowest
non-trivial order in the gauge coupling constant \cite{BaSc90},
to all orders  in other couplings in the model.  And we have already
proved the theorem to this order for the process $e^+e^- \to
Z_{\rm long}\gamma$. The diagram discussed by Donoghue and Tandean
\cite{DoTa93} is higher order in gauge coupling constant and hence is
not relevant for the validity of the equivalence theorem for the
process $e^+e^-\to Z\gamma$.

\begin{figure}
\large
\begin{center}
\begin{picture}(85,36)(-5,32)
\thicklines
\put(15,50){\line(-1,1){15}}     \put(7,58){\vector(1,-1){2}}
\put(15,50){\line(-1,-1){15}}      \put(7,42){\vector(1,1){2}}
\multiput(16,50)(4,0){7}{\oval(2,2)[t]}
\multiput(18,50)(4,0){7}{\oval(2,2)[b]}
\put(30,48){\makebox(0,0){\vector(-1,0){5}}}
\put(30,52){\makebox(0,0)[b]{$Z^\beta(K)$}}
\put(43,50){\line(1,1){12}}
\put(43,50){\line(1,-1){12}}
\put(55,38){\line(0,1){24}}
\put(55,50){\makebox(0,0){\vector(0,1){5}}}
\multiput(55,62)(3,0){10}{\line(1,0){2}}
\put(70,60){\makebox(0,0)[t]{$\pi^0(q)$}}
\put(83,62){\vector(1,0){3}}
\multiput(56,38)(4,0){7}{\oval(2,2)[t]}
\multiput(58,38)(4,0){7}{\oval(2,2)[b]}
\put(74,42){\makebox(0,0){\vector(1,0){5}}}
\put(74,42){\makebox(0,0)[b]{$\gamma^\alpha(k)$}}
\end{picture}
\end{center}
\caption[]{\sf Fermion-antifermion pair scattering into $\pi^0\gamma$
through anomalous triangle loops. There is another diagram where the
position of the $\pi^0$ and the photon are interchanged.}
\label{f:triangle}
	\end{figure}
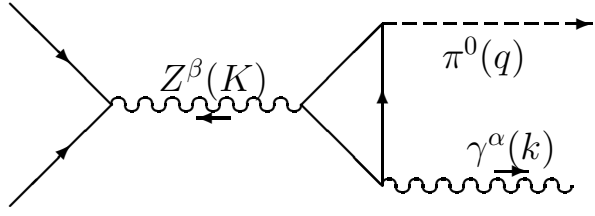

One can of course consider some other process for which
tree diagrams do not exist \cite{prcom}. Take, for example, the
process $\nu\bar\nu \to Z_{\rm long}\gamma$. Here, even the
non-triangle diagrams are fourth order in gauge coupling constants,
and so is the triangle-mediated diagram shown in Fig.\
\ref{f:triangle}. To examine the validity of the equivalence theorem
for this process, we will have to get into a detailed analysis
of the triangle part of the diagram. This has been done in earlier
papers \cite{DPO90,PhPh90,Hik90} in a different context. We mainly
follow the notation and analysis of Hikasa \cite{Hik90} in what follows.

The triangle part gives a $Z\gamma\pi$ effective coupling. Since only
the vector part of the $Z$ coupling matters, we can consider it as a
$\gamma^*\gamma\pi$ coupling apart from some irrelevant differences in
the coupling constants, $\gamma^*$ being a photon which is not
necessarily on shell. The matrix element for
$\pi^0\to \gamma^*\gamma$ transition is given by
	\begin{eqnarray}
\left< \left. \gamma^* (K) \gamma (k) \right| \pi^0 (q) \right> &=&
\lim_{q^2 \to m_\pi^2}  (m_\pi^2 - q^2)
\left< \gamma^* (K) \gamma (k) \left| \phi_\pi \right| 0 \right> \,.
\label{lsz}
	\end{eqnarray}
where $\phi_\pi$ is the interpolating field for the pion.
We now employ the PCAC relation:
	\begin{eqnarray}
\partial_\mu J_5^\mu = f_\pi m_\pi^2 \phi_\pi + {e^2 \over 16\pi^2}
F_{\mu\nu} \tilde F^{\mu\nu} \,,
\label{pcac}
	\end{eqnarray}
where $J_5^\mu$ is the axial vector current of the quarks, and
$F_{\mu\nu}$ is the electromagnetic field strength tensor. Using this,
we can rewrite the matrix element of Eq.\ (\ref{lsz}) as
	\begin{eqnarray}
\left< \gamma^* (K) \gamma (k) | \pi^0 (q) \right> &=&
{m_\pi^2 - q^2 \over f_\pi m_\pi^2}
\left\{ -e^2 \left< 0 \left| {\cal T} (\partial^\mu J_{5\mu} J_\alpha J_\beta)
\right| 0 \right>  - {e^2\over 4\pi^2} [kK]_{\alpha\beta} \right\}
\nonumber\\
&=& e^2 \, {m_\pi^2 - q^2 \over f_\pi m_\pi^2} \,
\left\{ q^\mu T_{\mu\alpha\beta} -
{1\over 4\pi^2} [kK]_{\alpha\beta} \right\} \,, \label{pigg}
	\end{eqnarray}
where $T_{\mu\alpha\beta}$ is the 3-point function with the axial
current and two vector currents, and
	\begin{eqnarray}
[kK]_{\alpha\beta}
&\equiv& \varepsilon_{\alpha\beta\lambda\rho}
k^\lambda K^\rho \,.
	\end{eqnarray}
Now, from the requirements of vector current conservation and
Lorentz invariance, one can write down the most general form for
$T_{\mu\alpha\beta}$ as follows:
	\begin{eqnarray}
T_{\mu\alpha\beta} &=& \phantom{+}
\left\{ k^2 \varepsilon_{\mu\alpha\beta\rho}
K^\rho + k_\alpha [kK]_{\mu\beta} \right\} F_1 \nonumber\\
&&+ \left\{ K^2 \varepsilon_{\mu\alpha\beta\rho}
k^\rho + K_\beta [kK]_{\mu\alpha} \right\} F_2 \nonumber\\
&&+ (k+K)_\mu [kK]_{\alpha\beta} F_3 + (k-K)_\mu [kK]_{\alpha\beta}
F_4\,,
	\end{eqnarray}
where $F_i\; (i=1\cdots4)$ are form factors\footnote{Note that $F_4=0$
from Bose symmetry. But we do not impose Bose symmetry at this level,
since we want to use our results in the case where the two vector
particles are the photon and the $Z$}.    Thus,
	\begin{eqnarray}
q^\mu T_{\mu\alpha\beta} &=&
\left\{ k^2 F_1 - K^2 F_2 - q^2 F_3 + (k^2-K^2) F_4 \right\}
[kK]_{\alpha\beta} \,.
\label{qT}
	\end{eqnarray}
We are obviously interested in the case $k^2=0$ since one photon is
on shell, and $q^2=0$ since the pion is a Goldstone boson in this
model. Thus, only the form factors $F_2$ and $F_4$ are relevant, and
from the general formulas given by Hikasa \cite{Hik90}, we obtain
	\begin{eqnarray}
F_2 &=& {1\over 2\pi^2} \int_0^1 dz \int_0^{1-z} dz'
{zz' \over m^2 - zz' K^2} \,,
\label{F2} \\
F_4 &=& 0 \,,
	\end{eqnarray}
where $m$ is the mass of the fermions in the loop. Now, the fermions
in the loop are obviously the $u$ and the $d$
quarks. Since the pion mass is zero, and $m_\pi^2 \propto m_u+m_d$,
the up and the down quarks must be massless in this model. Thus,
putting $m=0$ in Eq.\ (\ref{F2}), we obtain $F_2=-(4\pi^2K^2)^{-1}$,
so that from Eqs.\ (\ref{qT}) and (\ref{pigg}), we find that the
amplitude of the triangular loop actually vanishes in this case.

Two comments should be made here. First, the vanishing of this
amplitude has nothing to do with the decay $\pi^0\to \gamma\gamma$ in
the real world where the pion has a small mass
\cite{DPO90,PhPh90,Hik90}. For that case, in the soft pion limit one
is interested in the limit $k^2=K^2=0$ and $q^2$ small, so that the
form factor $F_3$ is important in that case. Moreover, since
$F_3=(48\pi^2m^2)^{-1}$, Eq.\ (\ref{qT}) shows that the term $q^\mu
T_{\mu\alpha\beta}$ vanishes in Eq.\ (\ref{pigg}) for $q^2=0$,
which is the famous Sutherland-Veltman theorem \cite{Sut67,Vel67}.
Second, in the present limit this implies that the contribution of
the diagram in Fig.\ \ref{f:triangle} is actually zero no matter which
fermion-antifermion pair is considered at the outer lines.  This is in
complete agreement with the equivalence theorem. The result can be,
and has been \cite{DPO90}, obtained by using a $\sigma$-model which
incorporates the anomalous contributions in a straightforward way.

We can summarize as follows. In models exhibiting dynamical symmetry
breaking, verification of the equivalence theorem may not be obvious,
but is nevertheless possible.  And we believe that the equivalence
theorem is always valid because it is based only on the requirement
that physical observables are continuous in the values of certain
parameters of the theory.

\paragraph*{Note added~:} After the paper was submitted for
publication, I was made aware of a paper by Zhang \cite{Zha89} which
also addresses the issue of the Equivalence theorem in models where
 symmetry is broken by fermion condensates. The conclusions of that paper
is similar to the present paper.

\paragraph*{Acknowledgements~:} The work was supported by the
Department of Energy, USA. I am indebted to D.~A. Dicus for long
discussions and to J.~F. Donoghue for  constructive criticism.
Conversations with D. Bowser-Chao,  A. El-Khadra, J. Gunion, F.
Olness, S. Weinberg and S. Willenbrock are also gratefully acknowledged.


\begin{thebibliography}{WW}

\bibitem{DoTa93} J.~F. Donoghue, J. Tandean: Phys. Lett. B301 (1993)
372.

\bibitem{CLT74} J.~M. Cornwall, D.~N. Leven, G. Tiktopoulos: Phys. Rev.
D10  (1974) 1145.

\bibitem{BQT77}
B.~W. Lee, C. Quigg, H.~B. Thacker: Phys. Rev. D16  (1977) 1519.

\bibitem{ChGa85}
M. Chanowitz, M.~K. Gaillard: Nucl. Phys.  B261 (1985) 379.

\bibitem{Wil88} S.~S.~D. Willenbrock: Ann. Phys. 186 (1988) 15.

\bibitem{BaSc90} J. Bagger, C. Schmidt: Phys. Rev. D41  (1990) 264.

\bibitem{Vel90} H. Veltman: Phys. Rev. D41 (1990) 2294.

\bibitem{Wei79} S. Weinberg: Phys. Rev. D19 (1979) 1277.

\bibitem{Sus79} L. Susskind: Phys. Rev. D20 (1979) 2619.

\bibitem{FLS72} K. Fujikawa, B.~W. Lee, R.~E. Shrock: Phys. Rev.
D13 (1972) 2674.

\bibitem{CBLP84} D. Chang, J. Basecq, L-F. Li, P.~B. Pal: Phys. Rev.
D30 (1984) 1601.

\bibitem{prcom} J. F. Donoghue: private communication.

\bibitem{DPO90} N. G. Deshpande, P. B. Pal, F. I. Olness: Phys. Lett.
B241 (1990) 119.

\bibitem{PhPh90} T. N. Pham, X. Y. Pham: Phys. Lett. B247 (1990) 438.

\bibitem{Hik90} K-I. Hikasa: Mod. Phys. Lett. A5 (1990) 1801.

\bibitem{Sut67} D. G. Sutherland: Nucl. Phys. B2 (1967) 433.

\bibitem{Vel67} M. Veltman: Proc. Roy. Soc. A301 (1967) 107.

\bibitem{Zha89} X. Zhang: Phys. Rev. D43 (1989) 3768.

	\end{thebibliography}
\end{document}